**Ultrafast Magneto-optical Fingerprints of Altermagnetism in MnTe**

Xu Yang[1,3,4,5], Xingkai Cheng[2,5], Zhuo Deng[1,3], Yu-Han Gao[1,3], Qing-Lin Yang[1,3], Zheng Chang[1,3], Peng-Tao Yang[1,3], Hong-Mei Feng[4], Xiang-Qun Zhang[1], Wei He[1], Junwei Liu[2,‡] and Zhao-Hua Cheng[1,3,4,‡]

[1]Beijing National Laboratory for Condensed Matter Physics, Institute of Physics, Chinese Academy of Sciences, Beijing 100190, China;
[2]Department of Physics, The Hong Kong University of Science and Technology, Hong Kong, China
[3]School of Physical Sciences, University of Chinese Academy of Sciences, Beijing 100049, China;
[4]Songshan Lake Materials Laboratory, Dongguan, Guangdong 523808, China.
[5]These authors contributed equally to this work.
[‡] **Corresponding authors:** zhcheng@iphy.ac.cn ; liuj@ust.hk

## Abstract

Recently identified altermagnets exhibit a distinctive dual-space nature: they possess spin-split electronic bands akin to ferromagnets in momentum space while maintaining the fully compensated magnetization of antiferromagnets in real space. This inherent duality, originating from the same crystal symmetry, gives rise to various intriguing physical phenomena unique to altermagnets. Consequently, a robust and efficient experimental signature capable of revealing this dual character is critically needed. The magneto-optical Kerr and Voigt effects, given their high sensitivity to ferromagnetism and antiferromagnetism, respectively, are ideally suited to probe this duality. Here, using time-resolved pump-probe magneto-optical measurements, we report the coexistence of pronounced Kerr and Voigt effects in the altermagnet MnTe. Combining the magnetization measurement and first-principles calculations, we demonstrate that the Kerr effect originates from the intrinsic Berry curvature of altermagnetism distribution in momentum space, while the Voigt effect arises from an anisotropic permittivity induced by the in-plane Néel order in real space, directly revealing the dual-space nature of altermagnets. Furthermore, the transient Kerr signal exhibits faster relaxation dynamics than the transient Voigt signal, underscoring their distinct origins in Berry curvature and Néel order, respectively. These findings establish transient magneto-optical responses as distinctive fingerprints of altermagnetism and position altermagnets as promising platforms for manipulating magneto-optical phenomena in ultrafast spin-optoelectronics.

## Introduction

Recently, a new class of spin-splitting anti-ferromagnets (AFMs), also named altermagnets (AMs), has been identified [1-4]. These materials combine ferromagnetic (FM)-like spin-split band structures with AFM-like vanishing net magnetization, thereby merging advantageous features of both FM and AFM systems [4-6]. However, up to date, experimental technique captures the dual nature simultaneously are extremely limited. For example, while angle-resolved photoemission spectroscopy (ARPES) directly measures momentum-resolved spin-split band structures [7-12], it cannot concurrently confirm antiparallel spin sublattices in real space. This inherent duality demands characterization in both real and reciprocal space. Consequently, there is a critical need for a robust and efficient experimental signature that can simultaneously capture this dual character. Magneto-optical (MO) effects are widely employed to probe the spin-related physical properties and electronic structures of materials [13-15]. In reflection geometry, two prominent MO effects are distinguished by their behavior under time-reversal symmetry ($T$) [16-19]: the $T$-odd magneto-optical Kerr effect (MOKE) and the $T$-even magneto-optical Voigt effect (MOVE). Consequently, the Kerr signal varies as an odd function of the applied magnetic field (or correspondingly, the magnetization in magnetic materials), while the Voigt signal varies as an even function.

Both MOKE and MOVE have been extensively investigated in novel magnetic systems. Traditionally, MOKE serves as a highly sensitive probe for characterizing ferromagnets (FMs) whose signal scales linearly with magnetization $\mathbf{M}$. Its application has been further expanded by recent discoveries, including the Berry-curvature-dominated Kerr effect in topological materials [20-23], photon-energy-dependent dynamics [24-28], and in multiferroic and inorganic materials [29-31]. In fully compensated antiferromagnets (AFMs), however, the net magnetization $\mathbf{M} = \mathbf{M_A} + \mathbf{M_B}$ vanishes due to the antiparallel alignment of magnetic sublattices, where $\mathbf{M_A}$ and $\mathbf{M_B}$ represent the magnetizations of magnetic sublattice A and B. Consequently, any MO effect that is linear (odd) in $\mathbf{M}$, such as MOKE, is suppressed due to destructive interference of signals from opposing sublattices [32]. This explains why AFMs such as CuMnAs [32] and CoO [33] exhibit negligible MOKE. The observable MO effects in such systems are therefore either quadratic (even) in magnetization [51] or transient linear MO effects accessible under ultrafast optical excitation [52]. Based on this symmetry distinction, the *T*-even MOVE can serve as the primary diagnostic tool for characterizing Néel order in AFM materials [20,53,55]. Given the high sensitivity of MOKE to ferromagnetism and MOVE to antiferromagnetism, both effects are expected to be observable in altermagnets, thereby providing a potential fingerprint to identify this emerging magnetism.

Here, we demonstrate the coexistence of pronounced MOKE and MOVE in 30 nm-thick altermagnetic manganese telluride (MnTe) films, as investigated by time-resolved pump-probe measurements. By systematically varying the probe laser polarization angle, we successfully isolated distinct time-evolution curves governed by the MOKE and MOVE, respectively. Combining the magnetization measurement with first-principles calculations, we show that the strong MOKE signal in MnTe originates from Berry curvature of altermagnet MnTe distribution in momentum space, while the MOVE signal arises from anisotropic permittivity induced by in-plane Néel order in real space, directly capturing the dual nature of altermagnets. This distinction further manifests as a faster timescale in the MOKE-dominated process compared to its MOVE counterpart, consistent with the faster response of the electronic system (Berry curvature) relative to the spin system (Néel order). Our work directly reveals the dual nature of altermagnet MnTe via the MO effect and provides the coexistence of MOKE and MOVE as a fingerprint of altermagnets, establishing altermagnets as a novel platform for investigating MO mechanisms, with particular relevance for ultrafast spintronic and spin-optoelectronic devices.

## Results and Discussion

We start from the definition of MOKE and the Voigt signal. The Kerr rotation $\theta_K$ and ellipticity $\eta_K$ in polar magneto-optical geometry can be written as (details see Supplemental Material S1[34]) [13,15,31],

$$\Phi_K = \theta_K + i\eta_K \approx -\frac{\epsilon_{xy}}{(1-\epsilon_{xx})\sqrt{\epsilon_{xx}}} = \frac{\sigma_{xy}(-H_z)}{\sigma_{xx}\left(1+\frac{i}{\epsilon_o\omega}\sigma_{xx}\right)^{\frac{1}{2}}}, \quad (1)$$

and the Voigt rotation $\theta_V$ and ellipticity $\eta_V$ are given by [15,17,19] ,

$$\Phi_V = \theta_V + i\eta_V = \frac{\epsilon_{xx}-\epsilon_{yy}}{\sqrt{\epsilon_{xx}}+\sqrt{\epsilon_{yy}}} = \frac{i}{\epsilon_0\omega}\frac{\sigma_{xx}-\sigma_{yy}}{\left(1+\frac{i}{\epsilon_o\omega}\sigma_{xx}\right)^{\frac{1}{2}}+\left(1+\frac{i}{\epsilon_o\omega}\sigma_{yy}\right)^{\frac{1}{2}}}, \quad (2)$$

where the $\epsilon_0$ is the permittivity of vacuum, and the $\omega$ is the frequency of light, and $\boldsymbol{\epsilon} = \boldsymbol{I} + \frac{i}{\omega\epsilon_0}\boldsymbol{\sigma}$ and $\boldsymbol{\sigma}$ is the optical conductivity.

According to the Onsager relations, the diagonal components of the optical conductivity tensor are even functions of the magnetic field H, while the off-diagonal components are odd functions. This symmetry is expressed as $\sigma_{xx}(\omega, -H_z) = \sigma_{xx}(\omega, H_z)$ , $\sigma_{yy}(\omega, -H_z) = \sigma_{yy}(\omega, H_z)$ and $\sigma_{xy}(\omega, -H_z) = -\sigma_{xy}(\omega, H_z)$ [16-19]. Therefore, MOKE signals are antisymmetric with respect to the applied magnetic field **H** (or magnetization **M** in magnetic materials). In contrast, the measured MOVE signals, which depend on diagonal components, are symmetric with respect to the applied magnetic field **H** (or magnetization **M** in magnetic materials).

To explore the MO response in MnTe, we first analyze the dielectric tensors dictated by different magnetic structures (ferromagnetic, antiferromagnetic, and altermagnetic) with different magnetic space groups (MSGs) that share the same crystal structure of MnTe (shown in Table I). This comparative approach allows us to isolate the distinct contributions of each magnetic order to the resulting MO properties (the results shown in Table I, details see the Supplemental Material S2[34]). In typical ferromagnetic systems with isotropic in-plane dielectric tensors, the reflection matrix satisfies $r_{sp} = -r_{ps}$, yielding a MOKE [14]. By contrast, antiferromagnetic systems exhibit vanishing off-diagonal dielectric terms due to $PT$ (or $tT$) symmetry. However, while anisotropic optical conductivity ($\sigma_{xx} \neq \sigma_{yy}$) produces distinct *s*- and *p*-light reflection coefficients ($r_{ss} \neq r_{pp}$) [51]. The diagonal component of the optical conductivity in paramagnetism MnTe (space group $P6_3/mm'c'$) is identical ($\sigma_{xx} = \sigma_{yy}$) due to the $C_{3z}$ symmetry.

When the in-plane Néel order breaks the $C_{3z}$ symmetry (magnetic space group $Cm'c'm$), it induces both diagonal terms $\sigma_{xx} \neq \sigma_{yy}$, and antisymmetric off-diagonal elements $\sigma_{xy}(\sigma_{yx})$. Consequently, the altermagnet MnTe should exhibit coexisting MOKE and MOVE signals.

To detect the MO effect of MnTe, a thin film with a thickness of 30 nm was fabricated by molecular beam epitaxy (MBE). Details on sample preparation, structural and magnetic characterization, anomalous Hall effect (AHE) measurements, angle-resolved photoemission spectroscopy (ARPES), and density functional theory (DFT) calculations are provided in Supplemental Material S3–S6 [34]. The sample exhibits a Néel temperature $T_N$~307 K, and a clear AHE below $T_N$. The time-resolved magneto-optical (ΔMO) and the transient reflectivity (Δ*R/R*) responses were acquired using an all-optical pump-probe system, with the experimental geometry detailed in Figure 1(a). Both beams are incident normally to the film plane. The surface normal of the film is parallel to the [001] crystallographic direction and is defined as the *z*-axis. A 790 nm pump beam (red) drives ultrafast suppression of altermagnetic (AM) order, while a frequency-doubled 395 nm probe (blue) monitors transient ΔMO signals through a cross-polarized detection scheme (Details see Supplemental Material S3[34]). The probe laser polarization angle $\theta$ specifies the orientation between the linear polarization (LP) vector and the Néel vector $\mathbf{N} = \mathbf{M_A} - \mathbf{M_B}$, where $\theta = 0°$ indicates LP parallel to $\mathbf{N}$.

Figure 1(b) displays time-resolved MO responses of the MnTe film measured at 80 K and 300 K under identical probe polarization geometry ($\theta = 45°$). The pump beam, polarized along the *x*-direction, delivered a fluence of 1.88 mJ/cm$^2$. A magnetic field of *H*=18 kOe was applied along with the *z*-axis during measurements. At temperature *T* = 300 K, only oscillatory features are detected, likely attributed to lattice vibrations (phonons) in MnTe. The observed oscillation frequencies of 53 GHz and 125 GHz correspond to acoustic phonon modes in MnTe, as reported in previous studies[51,52,54]. These frequencies are similar to those measured in $RuO_2$ (~50 GHz and 200 GHz)[55], confirming their origin as coherent acoustic phonons. Below $T_N$ (*T* = 80 K), the response exhibits two distinct regimes: (1) A sub-3 ps ultrafast decay reminiscent of magnetic order disruption in conventional ferro- or anti-ferromagnetic systems, and (2) oscillatory components spectrally distinct from the 300 K phonon modes. The signal stabilizes to a quasi-equilibrium state between 3–20 ps, followed by gradual recovery beyond 20 ps.

To characterize the angular dependence of MO responses, we systematically varied the probe polarization angle $\theta$ while monitoring ΔMO dynamics. Representative time-domain traces in Figure 2(a) demonstrate pronounced anisotropy in ΔMO amplitude across different $\theta$. Quantitative analysis reveals a periodic angular dependence of peak amplitudes（$\Delta t = 3.0$ ps） (Figure 2(b), black circles), exhibiting a 180° periodicity. For comparison, measuring polarization-dependent transient reflectivity (ΔR/R, Supplemental Material S7[34]) exhibits no angular modulation (Figure 2(b)). In polar magneto-optical geometry, the MO signal exhibits the following polarization dependence (derivation in Supplemental Material S1[34]):

$$\Delta \mathrm{MO} \approx \frac{2r_{sp}}{r_{ss}+r_{pp}} + \frac{r_{ss}-r_{pp}}{r_{ss}+r_{pp}} \sin 2\theta \tag{3}$$

The first term in the right-hand side corresponds to the MOKE signal, which remains invariant to probe polarization in polar geometry. The second term represents the MOVE signal, exhibiting a $\sin 2\theta$ dependence on the probe's polarization direction. We use equation (3) to fit the peak amplitudes (shown in Figure 2(b)), and separate the contributions of the Voigt and Kerr effects to the MO effect. Similar results are obtained at other delay times before the signal recovers to its pre-pump baseline, for example, at Δt=20.75 ps (Supplemental Material S8 Figure S7[34]). This consistency demonstrates that the observed Voigt and Kerr effects are independent of the specific time delay chosen for the analysis.

Several studies have reported that MnTe has a multi-domain structure, with three equivalent Néel vectors separated by 120°[8,56]. Therefore, the mixed magnetic domains in MnTe can influence the MOKE and MOVE signals; in principle, the signals from the different domains would cancel. However, the typical domain size in MnTe is smaller than 10 μm[56], which is quite similar to the diameter of the optical spot used in our experiments (Supplement materials S3[34]). Moreover, the sample was cooled under field (field cooled), which can significantly increase the average domain size compared to zero field cooling. Thus, the net MOKE and MOVE response does not vanish.

To unambiguously separate the Kerr and Voigt contributions in the MO curves, we examine the field dependence of the MO signal by reversing the magnetic field. Based on the fundamental definition of MO effects, the Kerr (Voigt) effect displays odd (even) symmetry with respect to the magnetic field. We change the direction of the magnetic field from 18 kOe to -18 kOe at $\theta = 0°$ and $\theta = 45°$, respectively. Figure 2(c) shows the MO signal measured at $\theta = 0°$ (Upper panel) and $\theta = 45°$ (Lower panel) with the $\pm$ 18 kOe magnetic field, which exhibits odd (Upper panel) and even (Lower panel) symmetry with the magnetic field. The open symbols in Figure 2(d) presents the difference curves between the 18 kOe and -18 kOe variations with time ($\frac{1}{2}\left(\mathrm{MO}(\theta,\ +18\ \mathrm{kOe}) - \mathrm{MO}(\theta\ , -18\ \mathrm{kOe})\right)$), and the maximum value is approximately 0.15 mrad, corresponding to the constant Kerr at different polarization angle. The solid symbols in Figure 2(d) presents the summation curves between the 18 kOe and -18 kOe variations with time ($\frac{1}{2}\left(\mathrm{MO}(\theta,\ +18\ \mathrm{kOe}) + \mathrm{MO}(\theta\ , -18\ \mathrm{kOe})\right)$) corresponding to the MOVE. The giant difference between the $\theta = 0°$ and $\theta = 45°$ implies the MOVE can be modulated by the polarization angle. Therefore, both the pronounced Kerr effect and the Voigt effect can survive in the altermagnet MnTe. This behavior distinguishes MnTe from conventional antiferromagnets such as CuMnAs, CoO, and $Mn_3Sn$, where the response is dominated by the Voigt effects[32,33,57]. Although $Mn_3Sn$ does exhibit a weak Kerr signal due to ferroic ordering of a cluster magnetic octupole [23], its magnitude is an order of magnitude smaller than its Voigt signal [57]. Furthermore, according to the theory of the Voigt effect in antiferromagnets, the pump-pulse polarization should have no influence on the Voigt effect[32]. We changed the pump light polarization and found that all pump polarizations are independent of the measured MO signals (Supplemental Material S9[34]). The coexistence of strong, tunable Kerr and Voigt responses in MnTe represents a clear departure from the behavior of conventional ferromagnetic and antiferromagnetic materials.

While the observed Voigt effect in MnTe is consistent with its antiferromagnetic arrangement in real space, the concurrent detection of a pronounced Kerr effect is particularly noteworthy. This significant Kerr response is governed by the off-diagonal component $\sigma_{xy}$ of the optical conductivity tensor, analogous to the conductivity term responsible for the anomalous Hall effect, where $\sigma_{xy}$ arises from contributions of both magnetic moment and Berry curvature $\mathbf{\Omega_z}$. Given the negligible net magnetization in MnTe, the pronounced Kerr effect is primarily attributed to the Berry curvature. In addition, we measured the changes in magnetic moment with magnetic field, as well as the variation of the MOKE signal with magnetic field (shown in the supplemental materials, S10[34]). During the magnetic field sweeps, we observed open hysteresis loops in the MOKE signal, whereas the net magnetization exhibited no hysteresis and was not proportional to the MOKE signal. Therefore, this discrepancy confirms that the spontaneous MOKE signal originates from the altermagnetic order in MnTe, rather than from a macroscopic net magnetization. Moreover, the difference in lineshape between the time-dependent traces of the MO rotation angle, $\Delta\theta(t)$, and ellipticity, $\Delta\eta(t)$, can serve as a unique fingerprint of "electronic (nonmagnetic)" contributions, which are not directly proportional to the true demagnetization $\Delta M(t)$. We analyzed the normalized time-dependent rotation and ellipticity curves for both the MOKE and MOVE signals in MnTe (see Supplemental Material S11[34]). Based on our analysis of the optical conductivity tensor and the qualitative analysis in the study, we conclude that the distinct behavior of these signals indicates that the MOKE in MnTe is mainly dominated by Berry curvature, whereas the MOVE has a magnetic origin.

Figure 3(a) presents the calculated distribution of Berry curvature, $\mathbf{\Omega_z}(\boldsymbol{k_x}, \boldsymbol{k_y})$, from first-principles calculation, with the Néel order aligned along the [010] direction. The finite $\mathbf{\Omega_z}(\boldsymbol{k_x}, \boldsymbol{k_y})$ across the Brillouin zone gives rise to a non-zero off-diagonal optical conductivity $\sigma_{xy}$, as shown in Figure 3(b2). It is crucial to note that while fully compensated AMs exhibit vanishing off-diagonal optical conductivity $\sigma_{xy} = 0$ in the ground state due to the cancellation between opposite sublattices, they are fundamentally distinct from conventional AFMs. In AFMs, spin-degenerate bands lead to exactly zero Berry curvature throughout the reciprocal space. In contrast, AMs allow both spin-splitting and finite Berry curvature at general k-points. However, the symmetry *C'* is a magnetic point group (MPG) operation associated with the crystal structure (which enforces compensated net magnetization in real space), forcing a cancellation of contributions from symmetry-connected momentum points (see Supplemental Material S12[34]), resulting in a net $\sigma_{xy} = 0$ [10,58]. Unlike the classical proportionality $\sigma_{xy} \propto \mathbf{M}$ in ferromagnets, AMs can host a giant $\sigma_{xy}$ even for a negligible net magnetization $\mathbf{M}$. This is because AMs preserve large, k-resolved Berry curvature (which is strictly zero in AFMs), and any finite magnetization disrupts the symmetry-enforced cancellation, thereby activating a large MOKE signal with nearly zero net moment (shown in Supplemental Material S10[34]).

Figure 3(b1) and (b3) display the nonequivalent diagonal optical conductivity components $\sigma_{xx}$ and $\sigma_{yy}$. This anisotropy originates from broken $C_{3z}$ symmetry induced by the in-plane Néel order, and is responsible for the Voigt effect. In magnetic materials, such anisotropic diagonal conductivity is directly linked to anisotropic magnetoresistance (AMR), which in turn reflects the orientation of the magnetic moment or Néel order. The combination of an AFM-like Voigt effect and a symmetry-protected, Berry-curvature-induced Kerr effect establishes a unique magneto-optical fingerprint for the altermagnet MnTe. This dual response highlights the intrinsic twofold nature of altermagnetism, bridging phenomena associated with both antiferromagnetic and ferromagnetic orders.

By considering the influence of the substrate, such as the lattice mismatch between the MnTe and $Al_2O_3$ substrate, we calculated the strain-induced MO effect. The results (see Supplemental Material S13[34]) show that the Kerr and Voigt effects remain robust. Recent studies have suggested that the MO effect may originate from the quantum metric[59]. However, in our experiment, with the magnetic easy-axis oriented along the [010] direction, the contribution from the quantum metric to the magneto-optical effect is zero under the $M_xT$ symmetry. Figure 3(c) presents the calculated MOKE and MOVE signals using Eqs. (1) and (2). The solid and dashed lines correspond to the real and imaginary parts, respectively, representing the rotation and ellipticity. The results clearly demonstrate that both pronounced MOKE and MOVE responses coexist in the altermagnet MnTe.

As shown in Figure 2, the total MO response of MnTe comprises contributions from both the Kerr and Voigt effects. To investigate the material's dynamical behavior, we analyze the time evolution of these components (Figure 4(a)). After isolating the non-oscillatory background (detailed in Supplemental Material S14[34]), we extract characteristic time constants from the transient signals. The fitting results reveal that the Kerr and Voigt effects exhibit distinct relaxation dynamics. The Kerr response, characterized by a single exponential decay with a time constant of $\tau_1^K \approx 445$ fs, resembles a one-step demagnetization process. In contrast, the Voigt dynamics are best described by a two-step process, requiring two time constants: a faster component $\tau_1^V \approx 642$ fs and a slower component $\tau_2^V \approx 1.42$ ps. This difference suggests that the underlying mechanisms driving the Kerr and Voigt responses in MnTe operate on separate timescales and likely involve distinct microscopic processes.

Based on the discussion above, we confirm that the MOKE signal originates from the off-diagonal optical conductivity component $\sigma_{xy}$, which is dominated by the Berry curvature $\mathbf{\Omega}_z$ distribution in momentum space. In contrast, the MOVE signals are governed by the difference of the diagonal optical conductivity components $\sigma_{xx} - \sigma_{yy}$, which is directly linked to the Néel order and thus to the antiferromagnetic structure in real space. Our DFT calculations indicate that the Néel order stems primarily from Mn atoms, with the occupied Mn 3d states located approximately 4 eV below the Fermi level (Figure 4(b)). The pump laser (photon energy 1.55 eV) directly excites electronic states within 1.55 eV of the Fermi energy. This excitation induces an immediate redistribution of the total Berry curvature, driving the fast Kerr response. Moreover, the transient MOKE curves exhibiting the same time constant under varying magnetic fields provide further evidence that the MOKE signal originates from the altermagnetic structure itself, rather than from the tiny magnetic moment (shown in Supplemental Materials S15[34]). Subsequently, energy is transferred to the deeper-lying states through interactions between electrons near the Fermi surface and those in more tightly bound orbitals. This slower energy-transfer process underlies the delayed Voigt dynamics, consistent with the picture described in earlier work[60]. Recent time-resolved quadratic magneto-optical Kerr effect (TR-QMOKE) studies of MnTe spin dynamics have employed a four-temperature model to explain such slower relaxation processes [61]. It was proposed that the large probe photon energy can detect magnons deep in the valence band, leading to the observed longer timescales [61]. Thus, the faster Kerr dynamics compared to the Voigt response align with the intrinsic mechanisms of MnTe's MO response: modifications in the Berry curvature (an electronic property) occur on ultrafast timescales, while changes in the total magnetic moment (governed by spin dynamics) evolve more slowly.

**Conclusion**

We have demonstrated pronounced MOKE and substantial MOVE in the altermagnet MnTe. The strong Kerr response is primarily attributed to the symmetry-governed Berry curvature distribution in momentum space, while the Voigt effect arises from the anisotropic permittivity induced by the in-plane Néel order. This distinct combination provides a clear MO fingerprint for identifying altermagnets. Furthermore, the transient MOKE signal exhibits faster relaxation dynamics than the transient MOVE signal, underscoring the dual electronic and magnetic nature of altermagnetism. These findings advance the fundamental understanding of unconventional magneto-optical responses in compensated magnets and position altermagnets as a promising platform for manipulating light and spin in ultrafast optoelectronics.

**Acknowledgments**

This work was supported by the National Key Research Program of China (grant No. 2024YFA1408702, 2021YFA1401500, and 2022YFA1403302), Hong Kong Research Grants Council (Grant No. 16303821, 16306722, 16304523, and C6046-24G), the National Natural Sciences Foundation of China (grant nos. U22A20115, and 52031015), Project supported by the Young Scientists Fund of the National Natural Science Foundation of China (Grant no. 12404140).

**Conflict of Interest**

The authors declare no conflict of interest.

**Author Contributions**

X.Y. and X. K. C. contributed equally to this work. Z. H. C. is responsible for the project and conceived this study and the experiments. X. Y. grew the samples and performed ARPES and time-resolved pump-probe measurements. X. K. C and J. L. carried out the symmetry analysis of different magnetic orders, the symmetry of the magneto-optical effect, and the first-principles calculations. Z. D, Y. H. G, H. M. F, Q. L. Y., and Z. C, help to measure x-ray diffraction, time-resolved pump-probe, and ARPES. H. M. F. made significant contributions to the setup of time-resolved Magneto-optical and transient reflectivity. X. Q. Z. and W. H. made significant contributions to the setup of MBE and ARPES. Z. H. C., and X. Y. drafted the manuscript. All authors discussed the results, worked on data analysis, and finalized the manuscript.

**Table I. Magneto-optical effects in ferromagnetic, antiferromagnetic, and altermagnetic systems.** The magneto-optical responses in ferromagnetic, antiferromagnetic, and altermagnetic systems arise from their respective permittivity tensors $\boldsymbol{\epsilon}$ and reflection matrices $\boldsymbol{r}$, which govern the emergence of magneto-optical phenomena.

| **Magnetic orders** | **FM** | **AFM** | **AM** | | |
|---|---|---|---|---|---|
| | **[001]** | **[010]** | **[100]** | **[001]** | **[010]** |
| **MSG** | $P6_3/mm'c'$ | $A_ama2$ | Cmcm | $P6_{3'}/m'm'c$ | $Cm'cm$ |
| **$\epsilon$ matrix from symmetry** | $\begin{pmatrix} \epsilon_{xx} & \epsilon_{xy} & 0 \\ -\epsilon_{xy} & \epsilon_{xx} & 0 \\ 0 & 0 & \epsilon_{zz} \end{pmatrix}$ | $\begin{pmatrix} \epsilon_{xx} & 0 & 0 \\ 0 & \epsilon_{yy} & 0 \\ 0 & 0 & \epsilon_{zz} \end{pmatrix}$ | $\begin{pmatrix} \epsilon_{xx} & 0 & 0 \\ 0 & \epsilon_{yy} & 0 \\ 0 & 0 & \epsilon_{zz} \end{pmatrix}$ | $\begin{pmatrix} \epsilon_{xx} & 0 & 0 \\ 0 & \epsilon_{xx} & 0 \\ 0 & 0 & \epsilon_{zz} \end{pmatrix}$ | $\begin{pmatrix} \epsilon_{xx} & \epsilon_{xy} & 0 \\ -\epsilon_{xy} & \epsilon_{yy} & 0 \\ 0 & 0 & \epsilon_{zz} \end{pmatrix}$ |
| **Reflection matrix $\boldsymbol{r}$ (xy plane)** | $\begin{bmatrix} r_{ss} & r_{sp} \\ -r_{sp} & r_{ss} \end{bmatrix}$ | $\begin{bmatrix} r_{ss} & 0 \\ 0 & r_{pp} \end{bmatrix}$ | $\begin{bmatrix} r_{ss} & 0 \\ 0 & r_{pp} \end{bmatrix}$ | $\begin{bmatrix} r_{ss} & 0 \\ 0 & r_{ss} \end{bmatrix}$ | $\begin{bmatrix} r_{ss} & r_{sp} \\ -r_{sp} & r_{pp} \end{bmatrix}$ |
| **MO Effect (xy plane)** | Kerr effect | Voigt effect | Voigt effect | No signal | Kerr and Voigt effect |

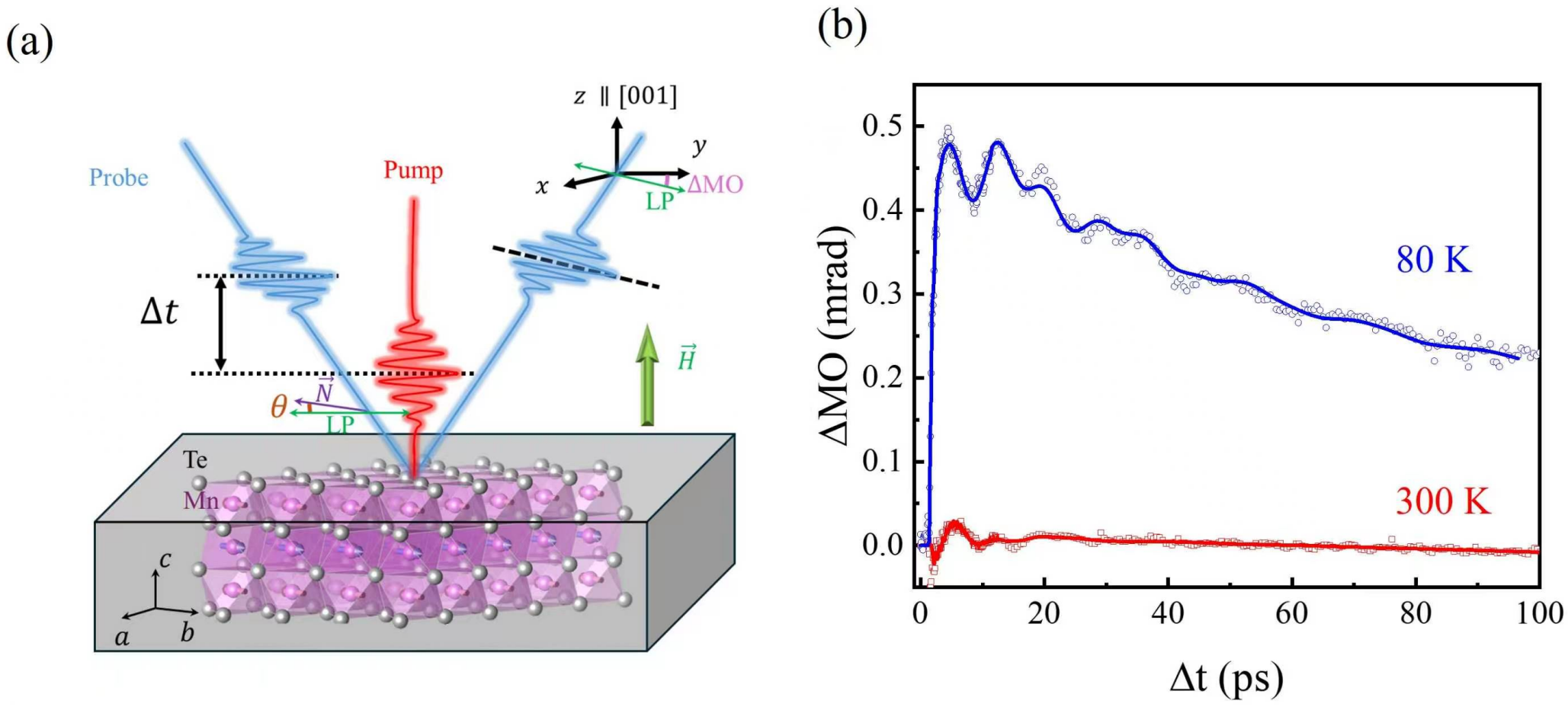

**Figure 1. Ultrafast magneto-optical response in MnTe. (a)** Schematic of the pump-probe setup. A 790 nm pump laser (red) induces rapid quenching of altermagnetic (AM) order in MnTe, while a 395 nm probe (blue) detects transient magneto-optical signals (ΔMO) via a cross-polarized configuration. The probe laser polarization angle $\theta$ is defined as the angular deviation between the linear polarization (LP) axis and the Néel vector $\vec{N}$, where $\theta = 0°$ corresponds to LP alignment parallel to $\vec{N}$. Both beams are incident perpendicular to the sample surface. A static field **H** is applied along the *z*-axis. Atomic structure: Pink spheres with pink or blue arrows depict Mn spin alignment, while gray spheres represent Te atoms. **(b)** Temperature-dependent ΔMO dynamics at $\theta = 45°$, contrasting responses at 80 K and 300 K, the pump fluence is 1.88 mJ/cm$^2$.

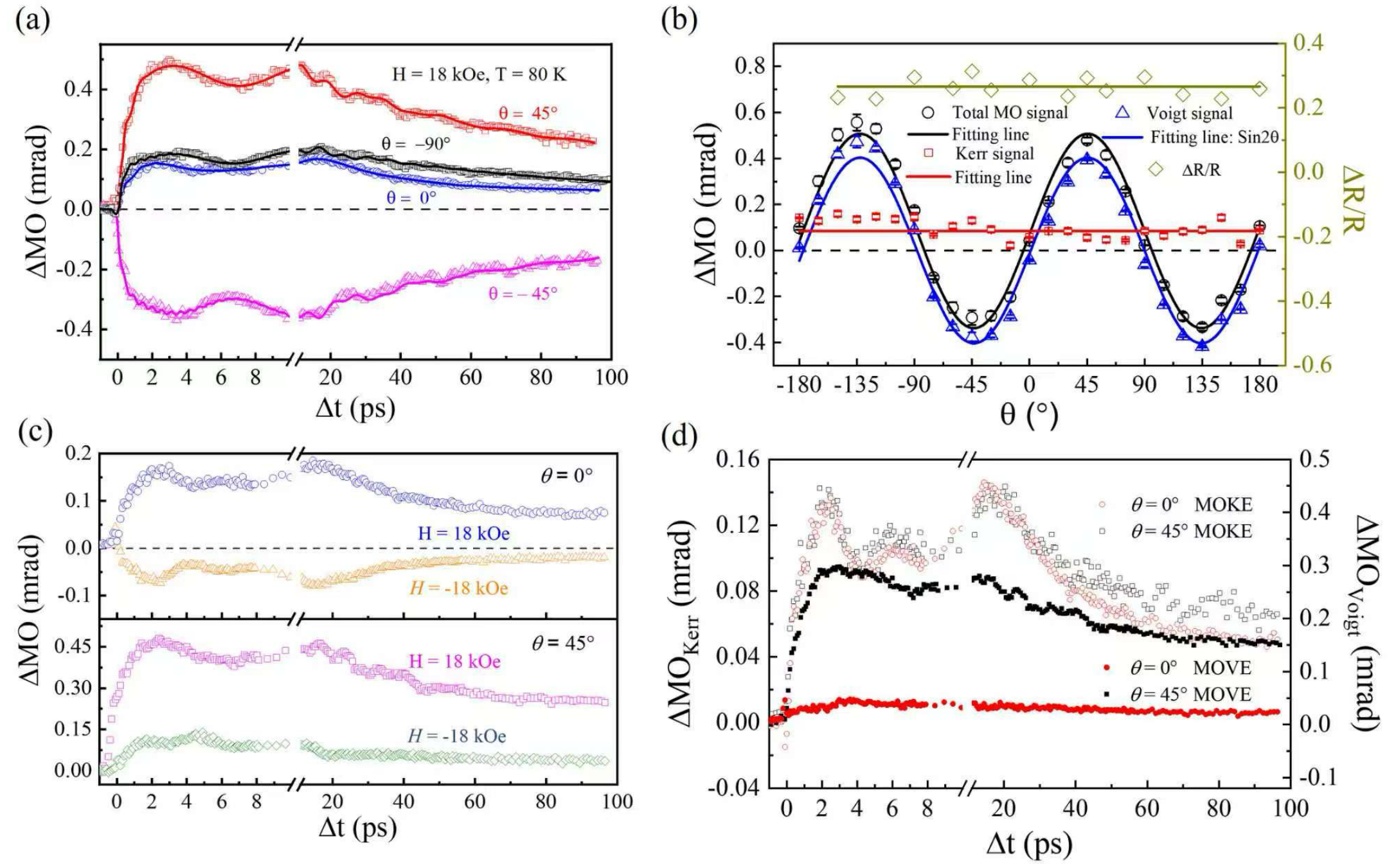

**Figure 2 Probe laser polarization dependence of the Voigt and Kerr signals. (a)** Time evolution of ΔMO at 80 K for different chosen probe laser polarizations $\theta$ (-90°, -45°, 0° and 45°) with the pump fluence is 1.88 mJ/cm$^2$. **(b)** The peak amplitude of the ΔMO ($\Delta t = 3.0$ ps) and ΔR/R signals was achieved from the best fitting results using equations (1) and (2). The black circle, blue triangle, and red squares represent the total ΔMO, the Voigt, and the Kerr signals. The brown diamonds represent the signal extracted from the ΔR/R. **(c)** The time-resolved MO curves measured at $\theta = 0°$ (Upper panel) and $\theta = 45°$ (Lower panel) at $H = \pm 18\,\mathrm{kOe}$. **(d)** The difference and summation between the $\pm 18\,\mathrm{kOe}$ curves at $\theta = 0°$ and $\theta = 45°$. The open symbols are the difference curves ( $\Delta\mathrm{MO_{diff}} = 1/2(\Delta\mathrm{MO}(\theta, H) - \Delta\mathrm{MO}(\theta, -H)$ ), which represent the MOKE. The solid symbols are the summation curves ($\Delta\mathrm{MO_{sum}} = 1/2(\Delta\mathrm{MO}(\theta, H) + \Delta\mathrm{MO}(\theta, -H)$), which represent the MOVE.

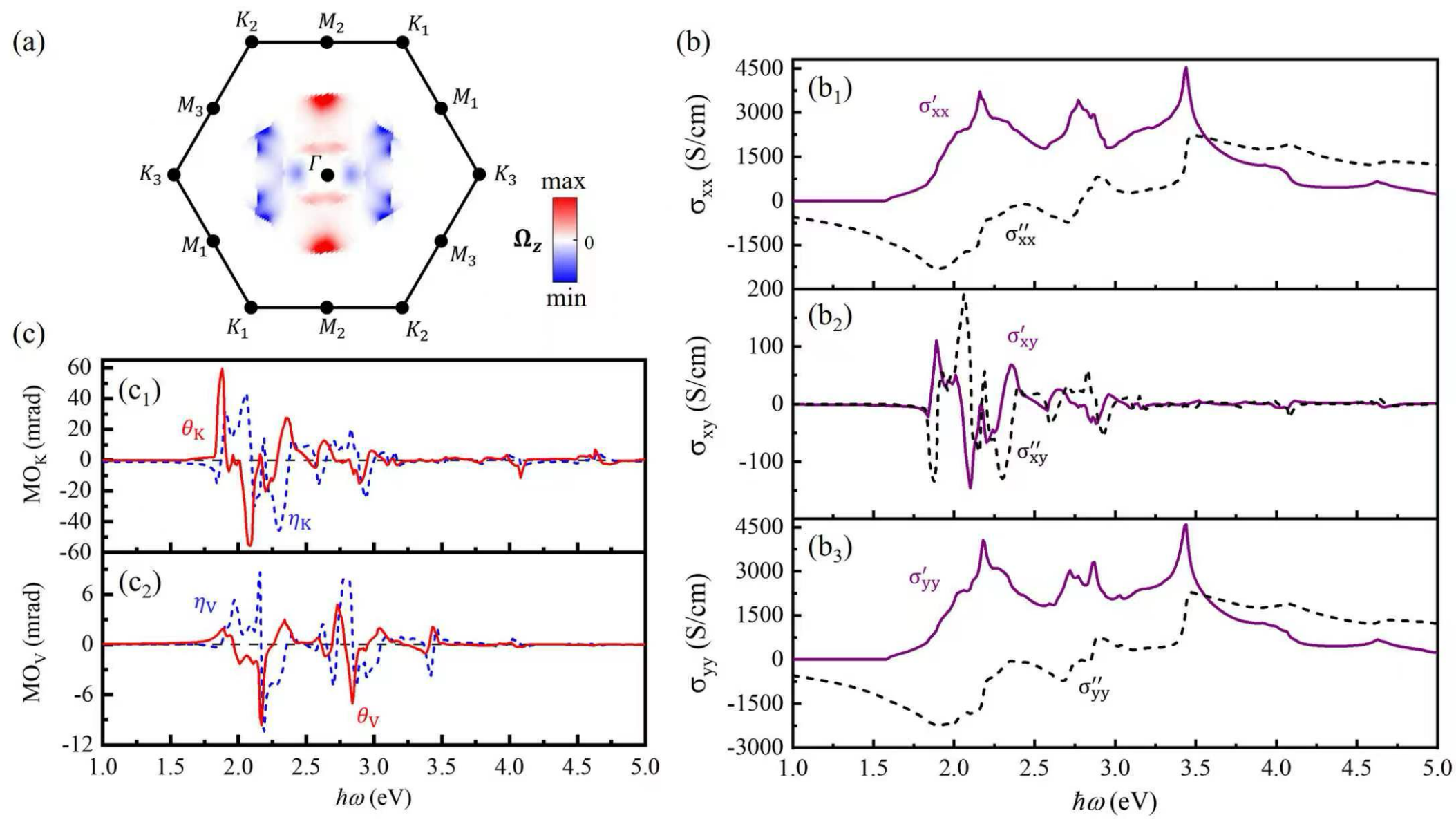

**Figure 3 The calculated k-dependence Berry curvature and magneto-optical spectra. (a)** Calculated k-dependence Berry curvature $\boldsymbol{\Omega_z}(\boldsymbol{k_x}, \boldsymbol{k_y}, 0)$ of MnTe. The blue and red colors represent the negative and positive Berry curvature at different k points. **(b)** The calculated optical conductivity tensor. **(b1)**, **(b2)**, and **(b3)** are $\sigma_{xx}$, $\sigma_{xy}$, and $\sigma_{yy}$. The solid and dashed lines represent the real part and the imaginary part of the optical conductivity. **(c)** Calculated the magneto-optical spectrum according to the optical conductivity tensor. **(c1)** and **(c2)** are the magneto-optical spectra of the Kerr effect and the Voigt effect, respectively. The solid and dashed lines represent the rotation angle and the ellipticity, respectively.

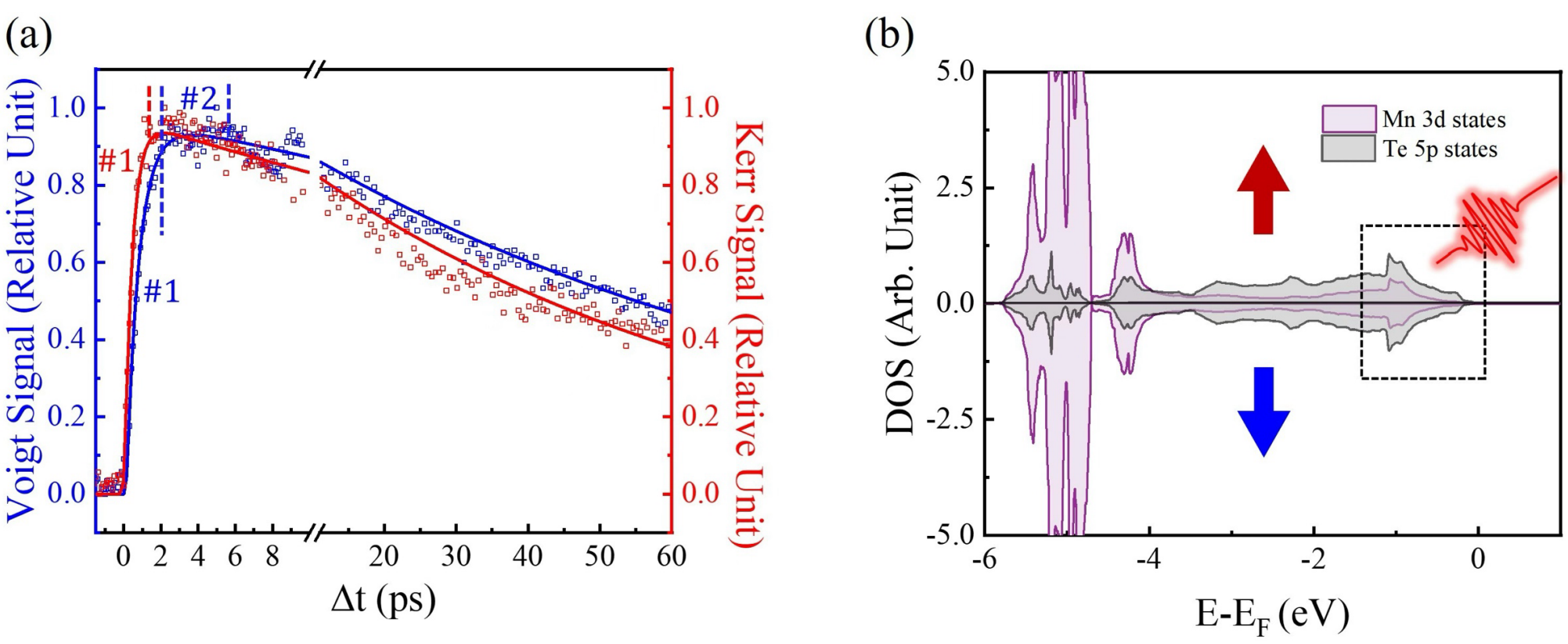

**Figure 4 Dynamic behavior of the MnTe. (a)** The dynamic curves of the Voigt effect and the Kerr effect. **(b)** the spin resolved density of states (DOS) of the MnTe. The red and blue arrows represent spin up and spin down, respectively.